\documentstyle{article}

\def\abstract#1{\vskip 7mm
        \begin{center}{\large Abstract}\par \smallskip
                \begin{minipage}[c]{12cm}
                        \small #1
                \end{minipage}
        \end{center}
}
\def\title#1{\begin{center}{\Large\bf #1}\end{center}}
\def\author#1{\vskip 5mm \begin{center}{#1}\end{center}}
\def\address#1{\begin{center}{\it #1}\end{center}}
\makeatletter
\@ifundefined{lesssim}{}{}
\@ifundefined{gtrsim}{}{}
\def\vereq#1#2{\lower3pt\vbox{\baselineskip1.5pt \lineskip1.5pt
\ialign{$\m@th#1\hfill##\hfil$\crcr#2\crcr\sim\crcr}}}
\makeatother

\begin{document}

\title{%
  Gravitational Waves in Brane World
  \smallskip \\
  {\large --- A Midi-superspace Approach ---}
}
\author{%
  Jiro Soda,\footnote{E-mail:jiro@phys.h.kyoto-u.ac.jp}
}
\address{
  Department of Fundamental Sciences, FIHS, Kyoto University, \\
  Sakyo-ku, Kyoto 606--8501, Japan
}
\abstract{
 It is important to reveal the brane-bulk correspondence for understanding
 the brane world cosmology. When  gravitational waves exist in the bulk,
  however, it is difficult to make the analysis of 
 the interrelationship between the brane and the bulk. Hence, the minimal
 model which allows  gravitational waves in the bulk would be useful.
 As for such models, we adopt the Bianchi type midi-superspace models. 
  In particular, the effects of  gravitational waves in the bulk on 
 the brane cosmology is examined using the midi-superspace approach. 
}

\section{Introduction}

 The idea of the brane world has provided an active area 
 in the field of cosmology. The main attractive point of this idea
 is its testability. However, to make a definite prediction,
 we need to understand the role of the bulk on the brane world cosmology.
 Before considering the complicated cosmological perturbation theory,
 we would like to try to understand the essence of the problem in the 
 simple models.  The aim of this paper is to study some aspects of  the  
brane-bulk correspondence in the Bianchi Type I midi-superspace model. 

 Let us start to describe the $Z_2$ symmetric brane world model 
 which we want to analyze from now on. The action for it
 is given by~\cite{RS,BWC5}
\begin{equation}
S= \frac{1}{2 \kappa^2}\int d^5 x \sqrt{-g}
\left(
{\cal R}^5 +  \frac{12}{l^2} \right)
- \sigma \int d^4 x \sqrt{-g_{brane}}
+ \int d^4 x \sqrt{-g_{brane}} {\cal L}_{matter}
\end{equation}
where $l$ denotes the curvature radius of the AdS spacetime, 
 $\kappa^2$ is the gravitational constant in the 5 dimensional spacetime,
 and $\sigma$ and  $g_{brane}$ represnt the brane tension and 
   the induced metric on the brane respectively. 
 Here we assume the relation
 $\kappa^2 \sigma=6/l$ which is necessary to have the Minkowski vacuum.

 From the above action, we obtain the 5-dimensional Einstein equations; 
\begin{equation}
G^M_N= \frac{6}{l^2} \delta^M_N + \kappa^2 
\frac{\sqrt{-g_{brane}}}{\sqrt{-g}} T^M_N  , 
\:\: (M,N =y,t,x^i) \ .
\end{equation}
As for the matter confined to the brane, we consider the perfect fluid
\begin{equation}
T^M_N= diag(0,-\rho,p,p,p) \delta(y)  \ .
\end{equation}

 First, let us consider the  FRW cosmology. 
The cosmological principle restricts the bulk metric
 in the following form:
\begin{equation}
ds^2 = e^{2 \beta(y,t)}(-dt^2+dy^2) 
    + e^{2 \alpha(y,t)}
\delta_{ij} dx^i dx^j    \ \ .
\end{equation}
Then, the induced metric on the brane becomes
\begin{equation}
ds^2 =-dt^2+e^{2 \alpha_0(t)} 
 \delta_{ij} dx^i dx^j    \ \ .
\end{equation}
 Due to the generalized Birkoff's theorem, 
 only AdS-Schwarzschild black hole solutions are allowed.
 Correspondingly, the cosmology on the brane is also simple.
 Putting  $8\pi G_4 =\kappa^2 /l$, we get the effective 
 Friedmann equation as
\begin{equation}
\dot{\alpha}_0^2 = \frac{8\pi G_4}{3} \rho + 
\frac{\kappa^4 \rho^2}{36} + e^{-4 \alpha_0} C_0  \ ,
\end{equation}
where the parameter $C_0$ corresponds to the mass of the black hole.
 It should be noted that  there exists no gravitational waves in this 
 back ground. This makes the understanding of  brane-bulk correspondence
 easier. 

 To proceed to the  cosmological perturbation theory is a natural next 
 step for understanding cosmology.~\cite{KJ,Muko,kodama,Bruck,Lan} 
 To reveal the nature of the cosmological perturbation, we need to consider
 the bulk spacetime with  gravitational waves.
 However, it is too complicated to analyze in detail. 

 In this circumstances,  we take the midi-superspace approach to attack 
 the issue. Namely, we consider the minimal model which can allow the bulk 
 gravitational waves. We shall consider the Bianchi type midi-superspace
 models:
\begin{equation}
ds^2 = e^{2 \beta(y,t)}(-dt^2+dy^2) 
   + e^{2 \alpha(y,t)}
                  g_{ij} (y,t) \omega^i \omega^j
   \ .
\end{equation}
Apparently, there can exist  (nonlinear)  gravitational waves
 in this bulk spacetime. The induced cosmology on the brane
 is nothing but the anisotropic bianchi type cosmology;
\begin{equation}
ds^2= -dt^2+ e^{2 \alpha_0 (t)}
                  g_{ij 0} (t) \omega^i \omega^j
                          \ .
\end{equation}
It includes FRW model as a special case. Hence, it should have implications
 into the cosmological perturbations.
 For simplicity, we will consider only the Bianchi type I model in this paper,
 althogh the generalization is straightforward.

\section{Anisotropic Cosmology}

Bianchi Type I midi-superspace metric is given by
\begin{equation}
  ds^2 = e^{2 \beta(y,t)}(-dt^2+dy^2) 
 + e^{2 \alpha(y,t)}
\left( e^{2(\chi_{+} (y,t) +\sqrt{3} \chi_{-} (y,t) )} dx_1^2 
+  e^{2(\chi_{+} (y,t) -\sqrt{3} \chi_{-} (y,t) )} dx_2^2 
+ e^{-4 \chi_{+} (y,t) } dx_3^2 \right)        \ .  
\end{equation}
 As is mentioned previously, gravitational waves can propagate in the bulk.
  From the above metric, it is easy to calculate the components of 
 the Einstein tensor relevant to the junction conditions as
\begin{eqnarray}
G^0_{\: 0} &=& - 3 e^{-2 \beta}(\dot{\alpha}^2 + \dot{\alpha} \dot{\beta} 
-\alpha''-2 \alpha'^2 + \alpha' \beta' -\dot{\chi}_{+}^2-\dot{\chi}_{-}^2
 -\chi^{\prime 2}_{+}-\chi^{\prime 2}_{-}) \\
G^{1}_{\:1} &=& 
 e^{-2 \beta}(-2 \ddot{\alpha} -3 \dot{\alpha}^2
- \ddot{\beta} + 2 \alpha''+ 3 \alpha'^2 + \beta''  
  +\ddot{\chi}_{+} +\sqrt{3} \ddot{\chi}_{-} -\chi''_{+}-\sqrt{3}\chi''_{-}
           \nonumber\\
&&   -3 \dot{\chi}_{+}^2 +3\chi^{\prime 2}_{+} 
          -3\dot{\chi}_{-}^2 +3\chi^{\prime 2}_{-}
   +3\dot{\alpha} \dot{\chi}_{+} -3\alpha' \chi'_{+}
   +3\sqrt{3}\dot{\alpha} \dot{\chi}_{-} -3\sqrt{3}\alpha' \chi'_{-}) \\
G^{2}_{\:2} &=& 
 e^{-2 \beta}(-2 \ddot{\alpha} -3 \dot{\alpha}^2
- \ddot{\beta} + 2 \alpha''+ 3 \alpha'^2 + \beta''  
   +\ddot{\chi}_{+} -\sqrt{3} \ddot{\chi}_{-} -\chi''_{+}+\sqrt{3}\chi''_{-}
                 \nonumber\\
&&   -3 \dot{\chi}_{+}^2 +3\chi^{\prime 2}_{+} 
           -3\dot{\chi}_{-}^2 +3\chi^{\prime 2}_{-}
   +3\dot{\alpha} \dot{\chi}_{+} -3\alpha' \chi'_{+}
   -3\sqrt{3}\dot{\alpha} \dot{\chi}_{-} +3\sqrt{3}\alpha' \chi'_{-}) \\
G^{3}_{\:3} &=& 
 e^{-2 \beta}(-2 \ddot{\alpha} -3 \dot{\alpha}^2
- \ddot{\beta} + 2 \alpha''+ 3 \alpha'^2 + \beta''  
    -2\ddot{\chi}_{+}  +2\chi''_{+} 
                    \nonumber\\
&&   -3 \dot{\chi}_{+}^2 +3\chi^{\prime 2}_{+} 
          -3\dot{\chi}_{-}^2 +3\chi^{\prime 2}_{-}
   -6\dot{\alpha} \dot{\chi}_{+} +6\alpha' \chi'_{+} ) \ .
\end{eqnarray}
The power series expansion near the brane  
\begin{equation}
\alpha(y,t)=\alpha_0(t)+ \alpha_1(t) \vert y \vert  + 
\frac{\alpha_2(t)}{2} y^2+ 
\cdot \cdot \cdot
\end{equation}
defines our notations.  From the Einstein equation,
 we get the junction conditions
\begin{eqnarray}
\alpha_1(t) &=& - \frac{1}{l}   
- \frac{\kappa^2 \rho(t)}{6} \ , \nonumber\\
\beta_1(t) &=&  - \frac{1}{l}+ 
\frac{\kappa^2 \rho(t)}{3} + \frac{ \kappa^2 p(t)}{2} 
                    \ ,  \nonumber\\
\chi_{+ 1}&=& 0   \ , \nonumber\\
\chi_{- 1} &=& 0  \ ,
\end{eqnarray}
where we set $e^{\beta_0(t)}=1$. 

The other components of the Einstein tensor
\begin{eqnarray}
G^y_{\: y} &=& 3 e^{-2 \beta}
( -\ddot{\alpha}-2 \dot{\alpha}^2 + \dot{\alpha} \dot{\beta}
+ \alpha'^2 + \alpha' \beta'
-\dot{\chi}_{+}^2-\dot{\chi}_{-}^2 -\chi^{\prime 2}_{+}-\chi^{\prime 2}_{-} ) 
  \\
G^0_{\: y} &=&  -3 e^{-2 \beta}
( \beta' \dot{\alpha}+ \alpha' \dot{\beta} - \dot{\alpha}'
-\dot{\alpha} \alpha' 
 -2\dot{\chi}_{+} \chi'_{+}  -2\dot{\chi}_{-} \chi'_{-}) 
\end{eqnarray}
give equations on the brane
\begin{eqnarray}
&& \ddot{\alpha}_0+2 \dot{\alpha}_0^2= \frac{\kappa^2}{2 l}
\left(\frac{\rho}{3}-p \right) - \frac{\kappa^4 \rho(\rho+3 p)}{36}
 -\dot{\chi}_{+}^2-\dot{\chi}_{-}^2 
\nonumber\\
&& \dot{\rho}+ 3 \dot{\alpha}_0 (\rho+p)=0  \ .
\nonumber
\end{eqnarray}
From these equations, we can deduce the Friedmann equation as
\begin{eqnarray}
  H^2 &=& {8\pi G_4 \over 3}\rho + \dot{\chi}_{+ 0}^2 + \dot{\chi}_{- 0}^2
    \quad   +{\kappa^4 \over 36} \rho^2 
  + C_0 e^{-4\alpha_0 } - e^{-4\alpha_0 } \int d\alpha_0 e^{-2\alpha_0}
    {d\over d\alpha_0} [e^{6\alpha_0}
             ( \dot{\chi}_{+ 0}^2 + \dot{\chi}_{- 0}^2 ) ]    \\
    &=& {8\pi G_4 \over 3}\rho + \dot{\chi}_{+ 0}^2 + \dot{\chi}_{- 0}^2
      +{\kappa^4 \over 36} \rho^2 
  + C_0 e^{-4\alpha_0 } - 2e^{-4\alpha_0 } \int dt e^{4\alpha_0}
    \left[ \dot{\chi}_{+ 0} \chi_{+ 2} 
 + \dot{\chi}_{- 0} \chi_{- 2}  \right]  \ ,
\end{eqnarray}
where we have used equations for $\chi_{\pm 0}$ to obtain the last line.

The last term represents the effect of gravitational waves in the bulk
 on the brane world cosmology. This correction term is apparently
 non-local and hence we need the information of the bulk.

\section{Vacuum Brane}

As we want to understand the effects of  gravitational waves
 on the brane cosmology, the existence of the matter is not
 essential. Hence, we consider the vacuum brane.

\subsection{Trivial bulk and boundary theory}

It is well known that there exist exact solutions in the form 
\begin{equation}
  ds^2 = ({l\over y})^2 \left[dy^2 
             + g_{\mu\nu} (x^\lambda ) dx^{\mu} dx^{\nu} \right]
 \ , 
\end{equation}
where $g_{\mu\nu}$ is the 4-d vacuum metric. All of the Bianchi type
 vacuum solutions can be elevated to the exact brane world solutions.

 In the case of the Bianchi type I model,
 we have the exact Kasner type solution~\cite{frolov}
\begin{equation}
  ds^2 =  ({l\over y})^2 \left( -dt^2+dy^2 + t^{2p_{1}} dx_1^2 
+  t^{2p_{2}} dx_2^2  + t^{2p_{3} } dx_3^2 \right)     \ .     
\end{equation}
Notice that the brane is located at $y=l$ in this coordinate system.
 As is usual, the parameters are constrained by the following relations
\begin{eqnarray}
  p_{1} + p_{2} + p_{3} &=& 1 \ , \\
  p_{1}^2 + p_{2}^2 + p_{3}^2 &=& 1 \ .
\end{eqnarray}
It would be interesting to consider the free scalar field on this background.
 As this system is separable, we can put
\begin{equation}
\phi = \chi(t) f(y) e^{ik_1 x_1+ik_2 x_2+ik_3 x_3}
\end{equation}
then the Klein-Gordon equation leads to
\begin{eqnarray}
 && \ddot{\chi}+{1\over t}\dot{\chi} +\omega^2 \chi +\left( k_1^2 t^{1-2p_1} 
 + k_2^2 t^{1-2p_2}+k_3^2 t^{1-2p_3} \right) \chi =0  \ , \\
 && -f'' + {3\over y} f' = \omega^2 f   \ .
\end{eqnarray}
Thus, we can define the boundary field theory as is done in AdS case.
 It is a generalization of the AdS/CFT  correspondence
 to  Gravity/Quantum Field Theory correspondence.

\subsection{ Effects of non-trivial bulk}

 Now we would like to consider how the cosmogical evolution
 on the brane is affected by the gravitational waves in the bulk.
 The strategy we take is the following. First we solve the equation
 on the brane which can be written using the variables solely on the
 brane.  We call this type of the equation as the constraint equation.
 Then we  calculate the non-local term from the explicit solution and
 identify its effective equation of state. 

The constraint equation on the brane is given by 
\begin{equation}
  \ddot{\alpha}_0 + 2\dot{\alpha}_0^2 
     +\dot{\chi}_{+ 0}^2 + \dot{\chi}_{- 0}^2 =0  \ .
\end{equation}
An interesting subclass of the solutions is
\begin{equation} 
\alpha_0 = A \log t  \ \ , \chi_{\pm 0} = B_{\pm} \log t 
    \ . 
\end{equation}
Substitution this ansatz into the constraint equation
 gives the relation: 
\begin{equation}
 2 (A -{1\over 4})^2 + B_{+}^2 + B_{-}^2 = {1\over 8} \ .
\end{equation}
Allowed region for $A$ becomes $  0 \leq A \leq {1\over 2} $.
The previous relations now become
\begin{eqnarray}
  p_{1} + p_{2} + p_{3} &=& 3A  \ , \\
  p_{1}^2 + p_{2}^2 + p_{3}^2 &=& 6A-9A^2  \ .
\end{eqnarray}
 In case $A=1/3$,  this reduces to exact Kasner type solution.
 For $B_{\pm} =0 , A= 1/2$, we obtain
 the  dark radiation dominated universe.

 From the above solutions, we get $\chi_{\pm 2}= (3A-1)B_{\pm}/t^2$.
 Using the previous formula, the effective Friedmann equation can be 
 deduced as
\begin{equation}
  H^2 =  \dot{\chi}_{+}^2 + \dot{\chi}_{-}^2  
                    + A(3A-1)  e^{-2\alpha_0 /A} \ . 
\end{equation}
 If $A>1/3$, the bulk effect accelerate the expansion.
The holographic projection of gravitational waves behaves
 as the fluid:
\begin{equation}
p=w \rho \ , \ \  w= {2\over 3A}-1 \ .
\end{equation}
In case $A<1/3$, the sound speed exceeds the velocity of light.
So it violates the holographic principle formulated by Fishler and Susskind.

\section{AdS/CFT correspondence}

 To analyze more general situations, we need a new apparatus.
 Here, we propose to use the AdS/CFT correspondence for that purpose. 
From AdS/CFT correspondence, we can deduce the low energy effective 
action for the brane world as
\begin{eqnarray}
  && S_{5-d EH} + \sigma \int d^4x \sqrt{g_{brane}} + S_{matter} 
     = 
      \frac{1}{16\pi G} \int d^4 x \sqrt{g_{brane}} R + S_{matter}
         \nonumber\\
  && \qquad + W_{CFT} 
         - \frac{l^3}{8\kappa^2} \log{\epsilon \over l} 
     \int d^4 x \sqrt{g_{brane}} \left[ R^{\mu\nu} R_{\mu\nu} -{1\over 3}R^2
               \right]  \ ,
\end{eqnarray}
where $\epsilon$ denotes the arbitraly parameter. 
In the  case of FRW cosmology, the last term vanishes like as
\begin{equation}
\int d^4 x \sqrt{g_{brane}} \left[ R^{\mu\nu} R_{\mu\nu} -{1\over 3}R^2
      \right] = 4 \int dt {d\over dt}[({\dot{a} \over N})^3] =0 \ .
\end{equation}
This result is consistent with the low energy effective Friedmann equation
\begin{equation}
\dot{\alpha}_0^2 = \frac{8\pi G_4}{3} \rho  
                       + e^{-4 \alpha_0} C_0  \ .
\end{equation}
 In contrast, for the Bianchi type I model, we have
\begin{eqnarray}
&&  \int d^4 x \sqrt{g_{brane}} \left[ R^{\mu\nu} R_{\mu\nu} -{1\over 3}R^2
               \right] \nonumber\\
  && \quad = 6\int dt e^{3\alpha} \left[
 (\ddot{\chi}_{+}+3\dot{\alpha} \dot{\chi}_{+})^2
  + (\ddot{\chi}_{-}+3\dot{\alpha} \dot{\chi}_{-})^2  
     +4(\dot{\chi}_{+}^2+\dot{\chi}_{-}^2)^2 
  +2(\ddot{\alpha} -\dot{\alpha}^2)(\dot{\chi}_{+}^2+\dot{\chi}_{-}^2) \right]
          \ .
\end{eqnarray}
 This does not vanish in general. 
Hence, the correction which we found in this work must be identified with
 this term. 
It would be interesting to analyze the effcts of the gravitational waves
 on the brane world cosmology using this correspondence.

\section{Conclusion}

 First we have suggested a possible extension of the AdS/CFT correspondence
 to Gravity/Quantum Field Theory correspondence in the case of the
 Kasner type exact solution
$$
  ds^2 =  ({l\over y})^2 \left( -dt^2+dy^2 + t^{2p_{1}} dx_1^2 
+  t^{2p_{2}} dx_2^2  + t^{2p_{3} } dx_3^2 \right)  \ .          
$$

 We have also shown that the bulk gravitational waves affect on the 
brane evolution as  a perfect fluid with equation of state, 
$ p=w \rho \ , \ \  w= {2/3A}-1 $. 
As $0 \leq A \leq 1/2$, we obtain $   w \geq 1/3 $. 
In particular, if $A\leq 1/3$,  the bulk effect is very stiff.
 Hence, it seems that the holographic principle is violated 
 in the brane world. It is also pointed out that the effects of  
gravitational waves 
 on the brane world cosmology can be investigated using the 
effective action derived from the AdS/CFT correspondence.

\end{document}